\documentstyle[aps,epsf,rotate]{revtex}
\begin{document}
\title{Realistic model of correlated disorder and Anderson localization}

\author{V.V. Flambaum$^1$, V.V. Sokolov$^2$}
\address{$^1$School of Physics, University of New South Wales,
Sydney 2052, Australia\\
$^2$Budker Institute of Nuclear physics, Novosibirsk 630090, Russia}
\maketitle

\date{\today}
\begin{abstract}

A conducting 1D chain or 2D film inside (or on the surface of) an insulator
is considered. Impurities displace the charges inside the insulator. This
results in a long-range fluctuating electric field acting on the conducting
line (plane). This field can be modeled by that of randomly distributed
electric dipoles. This model provides a random correlated potential with
$\langle U(r)U(r+k)\rangle \propto 1/k$. In the 1D case such correlations
may essentially influence the localization length but do not destroy Anderson
localization.


\end{abstract}



It was recently stated in \cite{Izrailev1998} that some special correlations
in a random potential can produce a mobility edge (between localized and
delocalized states) inside the allowed band in the 1D tight-binding model.
In principle, extrapolation of this result to 2D systems may give a possible
explanation of the insulator-conductor transition in dilute 2D electron
systems observed in ref. \cite{Kravchenko1996}.  In such a situation it is
very important to build a reasonable model of ``correlated disorder'' in
real systems and calculate the effects of this ``real'' disorder.

Usually, a 1D or 2D conductor is made inside or on the surface of an
insulating material. Impurities inside the insulator displace the electric
charges. However, a naive ``random charge'' model violates
electro-neutrality and gives wrong results. Indeed, the impurities do not
produce new charges, they only displace charges thus forming electric
dipoles. Therefore, we consider a model of randomly distributed electric
dipoles (alternatively, one can consider a spin glass model which gives the
correlated random magnetic field). The dipoles have long-range electric
field. Therefore, the potentials at different sites turn out to be
correlated.

The system of the dipoles $d_j$ produces the potential
\begin{equation}\label{U}
U(r)= e\sum_j {\bf d_j \nabla} \frac{1}{|\bf{r-R_j}|}.
\end{equation}
The average value of this potential is zero if $\langle {\bf d}_j\rangle =0$,
when the correlator of the potentials at the points ${\bf r_1}$ and ${\bf r_2}$ is
equal to \begin{equation}\label{UU} \langle U({\bf r_1}) U({\bf r_2})\rangle =
e^2\sum_{i,j} <{\bf d_i \nabla} \frac{1}{|\bf{r_1-R_i}|}
 {\bf d_j \nabla} \frac{1}{|\bf{r_2-R_j}|}>= \frac{e^2 {\bf d^2}}{3}
\sum_{j}\left({\bf \nabla}\frac{1}{|\bf{r_1-R_j}|}\right)
\left({\bf \nabla}\frac{1}{|\bf{r_2-R_j}|}\right) .
\end{equation}
Here we assumed that $\langle d_i^{\alpha}d_j^{\beta} \rangle ={\bf d}^2/3$
$ \delta_{il} \delta_{\alpha \beta}$ where $\alpha$ and $\beta$ are space
indices. Let us further suggest that the dipoles are distributed in space
with a constant density $\rho$. Then we have
\begin{equation}\label{UUrho}
\langle U({\bf r_1}) U({\bf r_2})\rangle
= \frac{e^2 {\bf d}^2 \rho}{3}
\int d^3 R\left({\bf \nabla} \frac{1}{|\bf{r_1-R}|}\right)
\left({\bf \nabla}\frac{1}{|\bf{r_2-R}|}\right)=
\frac{4\pi e^2 {\bf d}^2 \rho}{3|\bf{r_1-r_2}|} .
\end{equation}

However, the fluctuations of the potential at a given site ${\bf r_1=r_2}$
cannot be calculated in such a way since the expression (\ref{UUrho})
diverges.  To remove this divergence, one has to take into account the
geometrical size $r_0$ of the dipoles. Indeed, inside the radius $r_0$ the
electric field is not described by the dipole formula and the real potential
$U(r)$ does not contain the singularity $1/r^2$ which leads to the
divergence. Obviously, only those dipoles, which are closer to the
conducting chain than $r_0$, make the problem mentioned. One may eliminate
them by putting the chain into an empty tube with the diameter $d_0 > 2r_0$.
This method of cut-off is justified since a conducting wire always has finite
diameter.
The correlator (\ref{UUrho}) then reads on the chain oriented along the $x$
axis as
\begin{equation}\label{UUrhox}
\langle U(x_1) U(x_2)\rangle = \frac{e^2 {\bf d}^2 \rho}{3}
\frac{1}{\sqrt{(x_1-x_2)^2+d_0^2}}
{\bf E}\left[\frac{(x_1-x_2)^2}{(x_1-x_2)^2+d_0^2}\right]
\end{equation}
where ${\bf E}(m)$ is the complete second kind elliptic integral with
parameter $m$. This expression is everywhere finite and gives for the variance
$\langle U^2\rangle = (2\pi^2 e^2 {\bf d}^2 \rho)/d_0$ so that the normalized
correlator is equal to
\begin{equation}\label{cxi}
\xi(k)\equiv \frac{\langle U(x)U(x+k)\rangle}{\langle U^2\rangle}
 = \frac{2}{\pi}\frac{d_0}{\sqrt{k^2+d_0^2}}
{\bf E}\left(\frac{k^2}{k^2+d_0^2}\right).
\end{equation}
This function is positively definite, equal to 1 at $k=0$ and decays inversely
proportional to the distance $|k|$,
$$\xi(k) = \frac{2d_0}{\pi |k|},$$
when $|k|\gg d_0$. Similar calculation in the case of a conducting film which
is put in a split of the width $2z_0$ inside the insulator gives
\begin{equation}\label{fxi}
\xi({\bf k}) = \frac{\pi z_0}{|{\bf k}|}\left(1 -
\frac{2}{\pi}\arctan\frac{2z_0}{|{\bf k}|}\right).
\end{equation}

In the Ref. \cite{Thouless1972} the inverse localization length for electron
with an energy $E$, which moves in 1D random potential, has been expressed
in terms of the trace of the Green function. Being applied to the discrete
Schr\"odinger equation
\begin{equation}\label{psi}
\psi_{n+1}+\psi_{n-1}=(E+\epsilon_n)\psi_n
\end{equation}
with a weak potential $\epsilon_n$, the Thouless's formula looks in the second
order approximation like
\begin{equation} \label{l}
l^{-1}(E) =
-\frac{1}{2N} {\rm Re}\,\langle{\rm Tr}\left[G^{(0)}(E+i0)\,\epsilon\,
G^{(0)}(E+i0)\,\epsilon\,\right]\rangle.
\end{equation}
Here $N$ is the number of sites, $G^{(0)}(E)$ is the unperturbed Green
function of the equation (\ref{psi}), and $\epsilon$ is the diagonal matrix
of the random potential.

We employ spectral representation of the Green function $G^{(0)}(E)$
\begin{equation}\label{spG}
G^{(0)}_{n'n} = \frac{1}{N}\,\sum_q\frac{e^{i\mu_q(n-n')}}{E-E_q}
\end{equation}
to calculate the trace in the r.h.s of eq. (\ref{l}). Here
\begin{equation}\label{E,q}
E_q=2\cos\mu_q;\;\;\;\;\;\mu_q=\frac{\pi}{N}\,q;\;\;\;\;\;-N<q<N
\end{equation}
and the periodic boundary conditions are implied. Substitution in
eq. (\ref{l}) gives
\begin{equation}\label{lint}
l^{-1}(E) =- \lim_{N\rightarrow\infty}\frac{1}{2N}{\rm Re}\sum_{q,q'}
\left[\frac{1}{N^2}\sum_{n,n'}e^{i(\mu_q-\mu_{q'})(n-n')}
\langle\epsilon_n\,\epsilon_{n'}\rangle\right]
\frac{1}{(E+i0-E_{q'})(E+i0-E_{q}) }.
\end{equation}
In the limit $N\rightarrow\infty$ the summation over  $q$ and $q'$ can be
replaced by the integration over  $\mu_q$ and $\mu_{q'}$. After the change
of the variables $p=\mu_q-\mu_{q'}, k=n-n'$ the integrals and one of the sums
can be easily calculated. As a result, the inverse
correlation length $l^{-1}(E)$ at the energy $E=2\cos\mu$
\begin{equation}\label{lphi}
l^{-1}=\frac{\epsilon_0^2}{8 \sin^2 \mu}\,\phi(\mu)
\end{equation}
turns out to be proportional to the Fourier component
\begin{equation}\label{phimon}
\phi(\mu)= \sum_{k=-\infty}^{\infty}\,\xi(k)\,\exp(2i\mu k)
=\lim_{N\rightarrow\infty}\frac{1}{N\epsilon_0^2}\langle|
\sum_{n=-\infty}^{\infty}\epsilon_n\exp(2i\mu n)|^2\rangle
\end{equation}
of the normalized pair correlator
$\xi(k)=\langle\epsilon_n\epsilon_{n+k}\rangle/\epsilon_0^2$.  One can see
that the localization length is in essence the mean value of the squared
``Fourier component'' of the potential or in other words  the mean free path
for the backward scattering \cite{Thouless1972} (see also
\cite{Grin1988,Luck1989,Izrailev1998}) .

The Fourier transform of the correlator (\ref{cxi}) can be calculated
analytically in two limiting cases. If $d_0\ll 1$ (in the units of lattice
constant) the elliptic integral ${\bf E}(1)=1$ and
\begin{equation}\label{ll}
\phi(\mu) = 1-\frac{4d_0}{\pi}\ln|2\sin\mu|.
\end{equation}
It is less than one in the part $-\sqrt{3}<E<\sqrt{3}\;$ of the total energy
band $-2<E=2\cos\mu<2$ and is minimal, $\phi_{min}\simeq 1-0.9d_0$,
at the center of the spectrum $E=0,\; \mu=\pi/2$. The quantity $\phi_{min}$
reaches zero, which according to \cite{Izrailev1998} would mean delocalization, when
$d_0\simeq 1.1$. However, eq. (\ref{ll}) is not already entirely valid for such
values of $d_0$ and further terms in the expansion over $d_0$ must be taken into
account. In the opposite case of large $d_0$, transition to limit
$d_0\rightarrow\infty$ may be for all $\mu\neq 0,\pi$ taken before summing since the
resulting sum is cut-off by oscillations. Consecutive summation gives then
$\phi(\mu\neq 0,\pi)=0$.

For arbitrary finite values of $d_0$ Fourier components $\phi(\mu)$ remain
positive. However, they may be rather small within some energy interval.  In
Fig. 1 we show how $\phi(\pi/2)$ decreases with $d_0$ growing. It is seen
that at this point (and, obviously, in some vicinity of this point)
$\phi(\mu)$  becomes very small already for the values $d_0\simeq 3$
comparable with the lattice constant. In such a situation the deviations
from the tight-binding model and higher order corrections in $\epsilon_n$
may give important contributions and deserve a special consideration.

\begin{figure}
\unitlength 1cm
\begin{picture}(7,7)
\epsfxsize 12cm
\put(2,-3){ \rotate[r]{\epsfbox{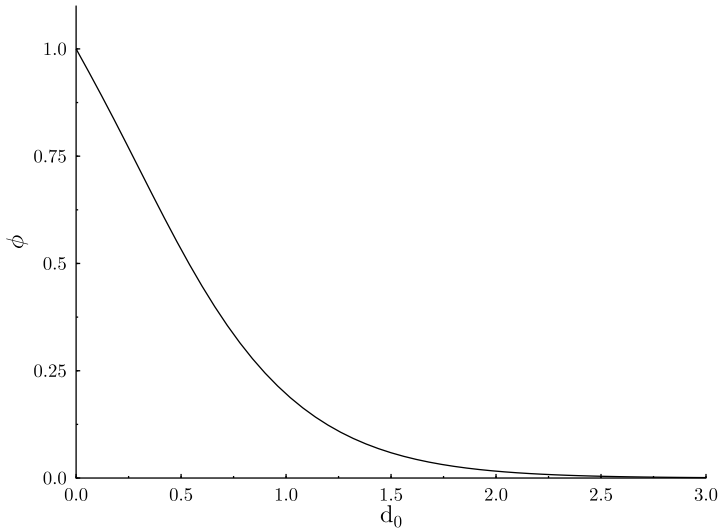}} }
\end{picture}
\caption{See text.}
\end{figure}

The realistic value of $d_0$ can hardly be much larger than the lattice
constant. Actually, it is a typical minimal distance from impurities to the
chain. Besides, any short-range fluctuations increase the variance
$\epsilon_0^2$ and reduce the normalized long-range correlator $\xi(k)$.
Thus, we conclude that the ``natural'' correlations in random potential can
hardly destroy Anderson localization in 1D case. However, these correlations
can significantly influence the value of the localization length. It is well
known that in 2D systems the localization length is very sensitive to
parameters of the problem. Therefore, in the 2D case the correlations due to
the long-range character of the dipole field may be even more important than
in the 1D case.

{\bf Acknowledgments.}
V.V. Flambaum acknowledges the support from Australian Research Council. He
is grateful to B. Altshuler, F. Izrailev and A. Krokhin for discussions and
to  V.Zelevinsky for valuable comments and hospitality during the stay in
MSU Cyclotron laboratory when this work was done. V.V. Sokolov is indebted
to V.F.~ Dmitriev, I.M. Izrailev and V.B. Telitsin for helpful discussions
and advices.

\end{document}